\documentclass{article}

\usepackage{graphicx}
\usepackage{bm}
\usepackage{amsthm,amsmath,amssymb}
\usepackage{mathrsfs}
\usepackage{hep,slashed}
\usepackage{indentfirst}
\usepackage[center]{titlesec}
\usepackage[title]{appendix}
\usepackage[colorlinks,linkcolor=red,anchorcolor=green,citecolor=blue]{hyperref}
\usepackage{diagbox,booktabs}
\usepackage{float}
\usepackage[a4paper,left=18mm,right=18mm,top=28mm,bottom=28mm]{geometry}
\usepackage{color,ulem}
\usepackage[T1]{fontenc}
\usepackage{authblk}

\begin{document}

\title{Leading and higher twist transverse momentum dependent parton distribution functions in the spectator model\thanks{Phys. Rev. D 104 (2021) 094043, \url{https://link.aps.org/doi/10.1103/PhysRevD.104.094043}}}
\author[1]{Xiao Liu}
\author[2]{Wenjuan Mao\footnote{Email address: wjmao@seu.edu.cn}}
\author[3]{Xiaoyu Wang\footnote{Email address: xiaoyuwang@zzu.edu.cn}}
\author[1,4,5]{Bo-Qiang Ma\footnote{Email address: mabq@pku.edu.cn}}
\affil[1]{School of Physics and State Key Laboratory of Nuclear Physics and Technology, Peking University, Beijing
100871, China}
\affil[2]{School of Physics and Telecommunication Engineering,
Zhoukou Normal University, Zhoukou 466000, Henan, China}
\affil[3]{School of Physics and Microelectronics, Zhengzhou University, Zhengzhou, Henan 450001, China}
\affil[4]{Center for High Energy Physics, Peking University, Beijing 100871, China}
\affil[5]{Collaborative Innovation Center of Quantum Matter, Beijing, China}

\renewcommand*{\Affilfont}{\small\it} 
\renewcommand\Authands{ and } 
\date{} 

\linespread{1.6}

\renewcommand\thesection{\Roman{section}}
\renewcommand\thesubsection{\arabic{subsection}}

\begin{sloppypar}

\maketitle

\begin{abstract}
    Transverse momentum dependent parton distribution functions, also abbreviated as TMDs, offer a three-dimensional picture of hadrons by taking the
    intrinsic transverse momentum of the parton into consideration. Hence, they are very important for us to understand the structure of hadrons.
    In this article, we calculate and summarize all TMDs of quark through the spectator model, from twist-2 to
    twist-4. Especially, we give complete results of TMDs at twist-4. We adopt a general analytical framework to calculate TMDs, with both scalar and axial-vector spectators being considered.
    All TMDs are calculated analytically in the light-cone coordinate, and single gluon rescattering is considered to
    generate T-odd TMDs. T-even TMDs are also calculated to this level, maybe for the first time. Different from the traditional point of view, the twist-4 TMDs
    can contribute to some physical observables like azimuthal asymmetries. An approximate formula of the Sivers asymmetry, including twist-4 TMDs, is given.\par
    \vspace{0.1cm}
\end{abstract}

\section{Introduction}

It is well known that the parton distribution function (PDF) in high energy particle physics is an important and useful tool to
study the properties of hadrons. PDFs can be extracted from experiments, and they provide information about deep structures of hadrons.
There are three PDFs --- unpolarized distribution function $f_1(x)$ (in the Amsterdam notation~\cite{MULDERS1996197}, the same below), helicity distribution function $g_1(x)$ and
transversity distribution function $h_1(x)$, and they have been extensively studied. These three distributions exhaust the information on the internal dynamics of the nucleon, if the
partons are assumed to be collinear with the parent nucleon.
Furthermore, if we take the transverse motion of partons into consideration, we can get transverse momentum dependent parton distribution functions (TMDs).
TMDs --- also called three-dimensional (3D) PDFs --- describe the probability density to find a parton in a hadron with longitudinal momentum fraction $x$
and transverse momentum $\bm{p}_T$ with respect to the parent hadron momentum~\cite{COLLINS1982445}. Unlike one-dimensional PDFs, TMDs give
a three-dimensional picture of the parton distribution in momentum space.\par

There is an increasing attention to TMDs in recent years, and a lot of theoretical and experimental researches have been
performed. For instance, TMDs can offer some contributions in several azimuthal spin asymmetries which were measured in semi-inclusive
deep inelastic scattering (SIDIS) and elsewhere~\cite{Yang:2018aue,Bacchetta_2007,Mao:2013waa,Mao:2014fma}. So, it is necessary to build a framework to calculate these TMDs. In this article,
we apply an analytical model --- the spectator diquark model --- to study all the TMDs~\cite{Jakob:1997wg}. As its name suggests,
the basic idea of the spectator diquark model~\cite{MR0421440} is to treat the intermediate states which are not collided directly by the
incoming particles (real or virtual) as spectator diquark states with effective masses. This model was originally proposed to study deep inelastic lepton nucleon scattering~\cite{CLOSE1973422,PhysRevD.15.2590}, based on the quark-parton model picture~\cite{PhysRevLett.23.1415,PhysRev.179.1547}.
For a proton target, the diquark can only be in spin 0 (scalar) state or spin 1 (axial-vector) state, due to the conservation of parity (the parities of the proton and the quark are both +1, so the parity of the diquark must be +1).
From a flavor point of view, the diquark could be an isoscalar ($ud$) or an isovector ($uu$). This idea is naive and intuitive but useful, and it can help us to understand some essential features of nucleon physical quantities including TMDs.\par

A particular realization of the diquark model picture in high energy physics is the light-cone quark-spectator-diquark model~\cite{MA1996320,MA1998461}, in which the relativistic effect of quark transverse motions is taken into account~\cite{Ma:1991xq,Ma:1992sj}.
This model plays an important role in investigating hadron structures by calculating relevant physical quantities, such as helicity distribution functions~\cite{MA1996320,CHEN2005188,Liu:2018mio}, transversity distribution functions~\cite{MA1998461}, form factors~\cite{PhysRevC.65.035205,PhysRevC.66.048201,PhysRevC.89.055202,PhysRevC.93.065209},
TMDs~\cite{LU2004200,PhysRevD.79.054008,PhysRevD.87.034037,LU2012451,PhysRevD.91.034019}, Wigner functions~\cite{PhysRevD.91.034019}, and so on.

In theoretical descriptions of high energy SIDIS processes involving hadrons, the cross sections are usually expanded in
power of $1/Q$, where $Q$ is the large momentum transfer of the collision. The cross sections in leading power can be expressed
as a convolution of the leading-twist (twist-2) distributions/fragmentation functions and the hard scattering coefficients~\cite{MULDERS1996197}.
In the subleading power, the twist-3 functions also contribute to the cross sections, and so on --- though one needs to consider the possibility that expressing cross sections in
 terms of parton distribution functions rests on factorization. However, it is worth mentioning that, there are some attempts to improve the availability of TMD factorization to subleading twist, such as
the work of Bacchetta \textit{et al.} in 2019~\cite{Bacchetta:2019qkv}, and a latest work in ~\cite{Vladimirov:2021hdn}.

In this article, we calculate and summarize all the
TMDs from twist-2 to twist-4. Certainly, we do all the calculations in the light-cone coordinate, which is a very natural framework
in high energy physics. To obtain nonzero results of T-odd TMDs, one has to consider the effect of the gauge links or the rescattering
between the struck quark and the spectator diquark~\cite{BRODSKY200299,JI200266,COLLINS200243}. Partial results of these TMDs at twist-2 and twist-3 have been given in earlier papers~\cite{Bacchetta:2008af,Jakob:1997wg,Metz2004,Yang:2018aue,Meissner:2007rx,Bacchetta2010,PhysRevD.81.114030},
we summarize them in a unified form. Other studies of twist-3 TMDs based on, e.g., light-front wave functions (LFWFs) can be found in Refs.~\cite{PASQUINI2019414,Rodini:2019ktv,Rodini:2019toi}. Notably, studies of twist-2 TMDs as overlaps of LFWFs have already been presented in detail in Ref.~\cite{Bacchetta:2008af}.
The contribution of gauge link to T-even TMDs is also considered, and the corresponding analytical results are calculated maybe for the first time.
We also give the results of TMDs at twist-4, which have not been completely calculated yet. Some studies involving unpolarized TMDs at twist-4 can be found in, e.g., Refs.~\cite{Lorce:2014hxa,PhysRevLett.67.552,SIGNAL1997415}.\par

The twist-4 TMDs are used to be regarded as pure theoretical quantities, and they are only of academic interest because of the unavailable factorization and
the lack of relations with observables. But, as pointed out by Zuo-tang Liang \textit{et al.} in Ref.~\cite{PhysRevD.95.074017}, one can obtain some concrete formulas
describing the contribution of twist-4 TMDs to several azimuthal asymmetries, if the cross section is expanded up to twist-4 in the framework of collinear expansion~\cite{ELLIS19821,ELLIS198329,QIU1991105,QIU1991137}.
Based on these formulas, we get an approximate expression for Sivers asymmetry including the contribution of twist-4 TMDs.

This article is organized as follows. In Sec. \ref{sec2}, we adopt an analytical framework to calculate TMDs.
We obtain two useful formulas for T-even and T-odd TMDs respectively. Then in Sec. \ref{sec3}, we give the complete
results of scalar-diqurak TMDs as well as axial-vector-diquark TMDs from twist-2 to twist-4 in the light-cone transversely polarized diquark states. Some
phenomenological discussions are also given there. In Sec. \ref{sec4} we give a twist-4 modified expression for the Sivers asymmetry.
We put the results of TMDs under another type of polarized diquark state and the results of T-even TMDs at one-loop level in the Appendix.

\section{Analytical framework}
\label{sec2}

In this section, we try to establish a framework, inspired by Ref.~\cite{Bacchetta:2008af}, to calculate TMDs analytically from the perspective of correlators, i.e., via the Dirac decomposition of the correlation function.
We put the results of these TMDs in a unified form in the next section and the Appendix.

As is showed in Ref.~\cite{Bacchetta:2008af}, the quark-quark correlation function in the quark-diquark model can be expressed as
\begin{equation}
    \label{correlator} \Phi(x,\bm{p}_T,S)=\left.\frac{1}{(2\pi)^3}\frac{1}{2(1-x)P^+}\overline{\mathcal{M}}^{(0)}(S)\mathcal{M}^{(0)}(S)\right|_{p^2=\tau(x,\bm{p}_T),p^+=xP^+},
\end{equation}
where $p$ is the four-momentum of the active quark, $\bm{p}_T$ is its transverse momentum, $m$ is its mass, and $x=p^{+}/P^{+}$ is the light-cone momentum fraction of the active quark to the hadron. The symbols $M$ and $M_X$ are the masses of the hadron and the spectator diquark, respectively.
This formula is also called the spectator approximation. The on-shell condition $(P-p)^2=M^2_X$ for the spectator diquark suggests that the quark is off shell:
\begin{equation}
    p^2\equiv\tau(x,\bm{p}_T)=-\frac{\bm{p}_T^2+L_X^2(m^2)}{1-x}+m^2,~\text{where}~L_X^2(m^2)=xM_X^2+(1-x)m^2-x(1-x)M^2. \label{L}
\end{equation}
Usually, we choose the frame where the hadron momentum $P$ has no transverse components~\cite{Bacchetta_2007,Bacchetta:2008af}. Thus, some four-vectors used in this
article can be written as ($\lambda_N$---we use $\lambda$ later for simplicity---is the helicity of the hadron),
$P=\left(P^+,\frac{M^2}{2P^+},\bm{0}\right)$,
$p=\left(xP^+,\frac{(1-x)M^2-\bm{p}_T^2-M_X^2}{2(1-x)P^+},\bm{p}_T\right)$,
$S=\left(\frac{\lambda_NP^+}{M},-\frac{\lambda_NM}{2P^+},\bm{S}_T\right)$,
satisfying $P\cdot S=0$, $S^2=-1$, i.e., $\lambda^2+\bm{S}_T^2=1$.

\begin{figure}[H]
    \centering
    \includegraphics[scale=0.6]{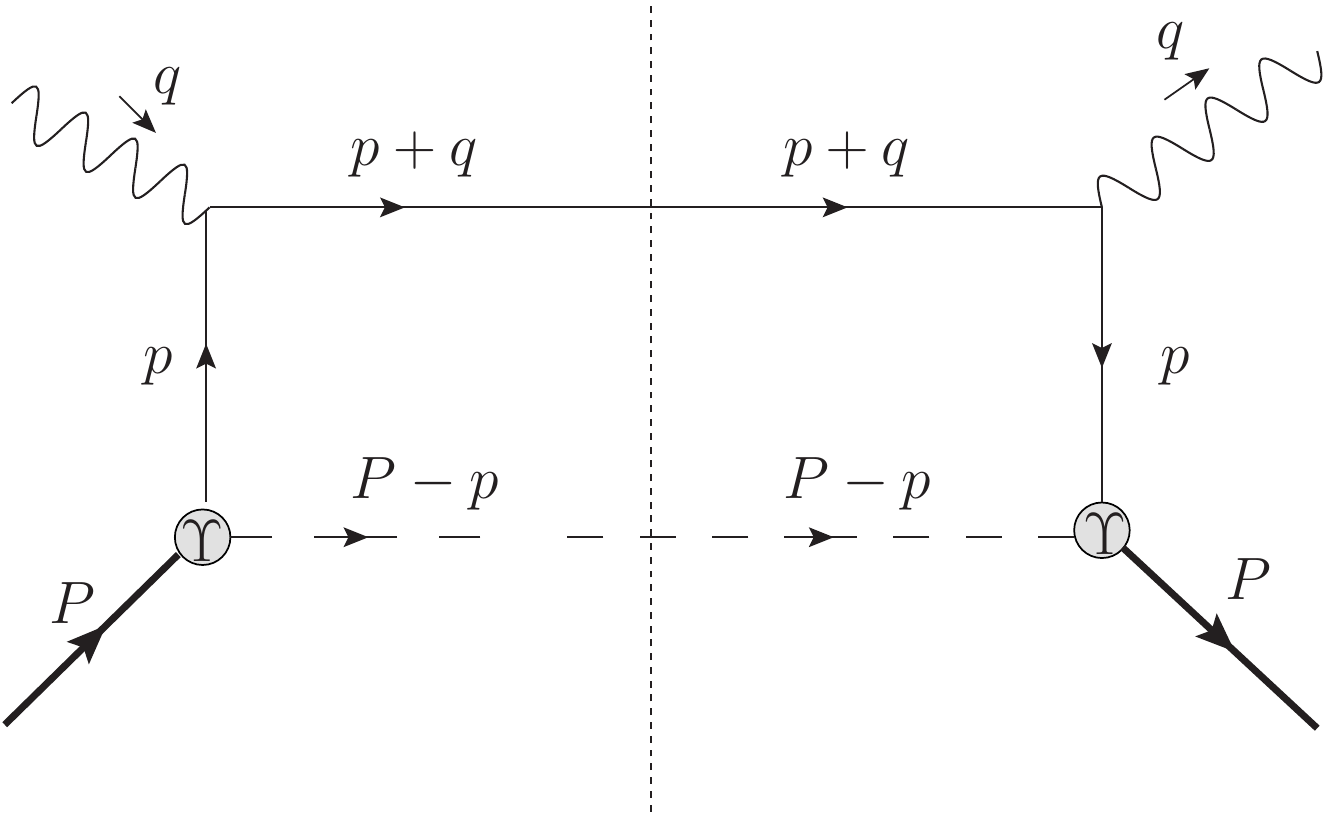}
    \caption{Tree level cut diagram for $\overline{\mathcal{M}}^{(0)}\mathcal{M}^{(0)}$. The dashed line denotes the diquark.}
    \label{fig1}
\end{figure}

The symbol $\mathcal{M}^{(0)}$ here is the nucleon-quark-diquark scattering amplitude at tree level (see the lower-left part of Fig.~\ref{fig1}) as follows:
\begin{equation}
    \mathcal{M}^{(0)}=\langle P-p|\Psi(0)|P\rangle=
\begin{cases}
    \frac{\bm{i}}{\slashed{p}-m}\Upsilon_sU(P,S) &\text{scalar diquark},\\
    \frac{\bm{i}}{\slashed{p}-m}\varepsilon^*_\mu(P-p,\lambda_a)\Upsilon_a^\mu U(P,S) &\text{axial-vector diquark},
\end{cases}
\end{equation}
and $\overline{\mathcal{M}}^{(0)}$ is its Hermitian conjugation multiplied by $\gamma_0$. Here $\varepsilon(P-p,\lambda_a)$ is the polarization vector of the axial-vector diquark with
momentum $P-p$ and helicity $\lambda_a$. There are several choices used when summing over all polarization states~\cite{Bacchetta:2008af}, and we set $d^{\mu\nu}=\sum_{\lambda_a}\varepsilon^*_\mu\varepsilon_\nu$,
\begin{equation}
  \label{eqd}
  d^{\mu\nu}(P-p)=
  \begin{cases}
    -g^{\mu\nu}+\frac{(P-p)^{\mu}n_-^{\nu}+(P-p)^{\nu}n_-^{\mu}}{(P-p)\cdot n_-}-\frac{M_a^2}{[(P-p)\cdot n_-]^2}n_-^{\mu}n_-^{\nu} &\text{(see Ref.~\cite{BRODSKY2001311})},\\
    -g^{\mu\nu}+\frac{(P-p)^{\mu}(P-p)^{\nu}}{M_a^2} &\text{(see Ref.~\cite{PhysRevD.77.094016})},\\
    -g^{\mu\nu}+\frac{P^{\mu}P^{\nu}}{M_a^2} &\text{(see Ref.~\cite{Jakob:1997wg})},\\
    -g^{\mu\nu} &\text{(see Ref.~\cite{BACCHETTA2004109})}.
  \end{cases}
\end{equation}
According to Ref.~\cite{Mao:2013waa}, only the first and the last choices of $d^{\mu\nu}$ can give convergent results of some TMDs, like $g^\perp$, even if the dipolar coupling factor (see below) is applied.
In this article, we give the results of all TMDs under the first and the last choice of $d^{\mu\nu}$ from twist-2 to twist-4. The results at twist-2 can also be found in Refs.~\cite{Jakob:1997wg,Meissner:2007rx,Bacchetta:2008af}.
The results of the other choices of $d^{\mu\nu}$ can also be calculated through this analytical framework.\par

In the above expression of nucleon-quark-diquark scattering amplitude, the symbol
$\Upsilon_{s/a}$ denotes the nucleon-quark-diquark vertex ($s$ for the scalar diquark and $a$ for the axial-vector diquark, the same below), and we choose the simplest forms of them~\cite{Jakob:1997wg} (other forms can be found in, e.g., Refs.~\cite{PhysRevD.77.094016,Jakob:1997wg}),
\begin{equation}
    \Upsilon_s(p^2)=g_s(p^2)\bm{1}(\text{identity matrix}),~~\Upsilon_a^\mu(p^2)=\frac{g_a(p^2)}{\sqrt{2}}\gamma^{\mu}\gamma^{5},
\end{equation}
here $g_X(p^2)$ is the coupling factor of the nucleon-quark-diquark vertex that takes into account the
composite structure of the nucleon and the diquark spectator. The most general structure of the vertices can be found in Ref.~\cite{PhysRevD.49.1183}. Usually, there are three common choices of $g_X(p^2)$,
\begin{equation}
    g_X(p^2)=
    \begin{cases}
        g_X^\text{p.l.} & \text{pointlike},\\
        g_X^\text{dip}\frac{p^2-m^2}{|p^2-\Lambda_X^2|^2} & \text{dipolar},\\
        g_X^\text{exp}e^{(p^2-m^2)/\Lambda_X^2} & \text{exponential},
    \end{cases}
\end{equation}
where $\Lambda_X$ is the cutoff parameter, and $g_X$ is the coupling constant. In order to regularize the light-cone divergence appearing
in the calculations of T-odd TMDs
~\cite{GAMBERG2006508,PhysRevD.77.114026} and avoid the end-points singularities~\cite{BURKARDT2002311}, the second choice, i.e., the dipolar form for $g_X(p^2)$ is usually used~\cite{Bacchetta:2008af,Yang:2018aue}, and it can be rewritten as follows:
\begin{equation}
    g_X(p^2)=g_X\frac{p^2-m^2}{|p^2-\Lambda_X^2|^2}=g_X\frac{(p^2-m^2)(1-x)^2}{(\bm{p}_T^2+L_X^2(\Lambda_X^2))^2}~(X=s~\text{or}~a),
\end{equation}
where  $L_X^2(\Lambda_X^2)$ ($L_X^2$ for simplicity) has the same form with Eq.~(\ref{L}) by replacing $m^2$ with $\Lambda^2_X$.
In this article, we give two formulas to calculate TMDs analytically under a general form of the coupling factor $g_X(p^2)$. Then, the concrete results of TMDs under the
dipolar coupling of $g_X(p^2)$ are shown. Besides, results of twist-2 TMDs under all these choices are already given in Ref.~\cite{Bacchetta:2008af}. \par

\begin{figure}[H]
    \centering
    \includegraphics[scale=0.6]{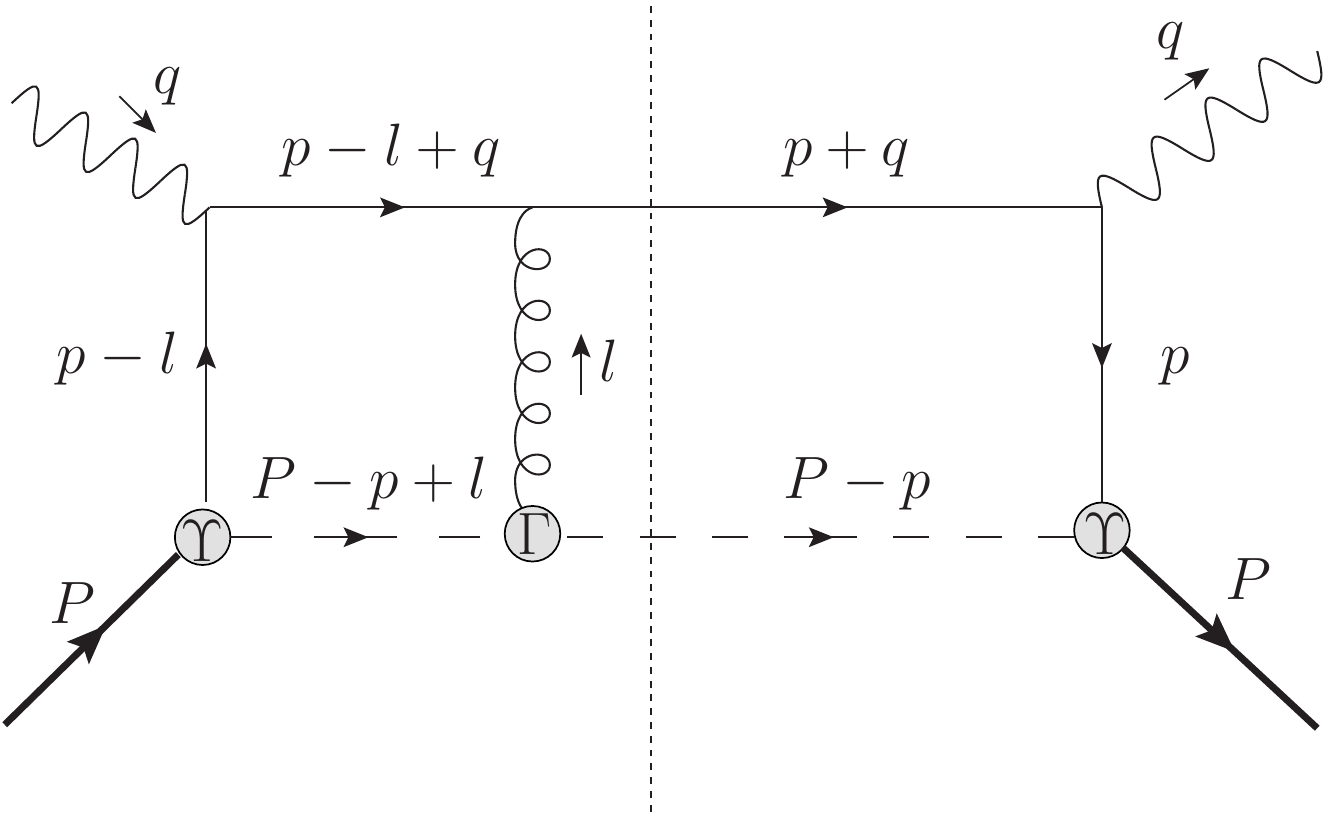}
    \caption{One-loop level cut diagram for $\overline{\mathcal{M}}^{(0)}\mathcal{M}^{(1)}$. The Hermitian conjugate diagram is not shown.}
    \label{fig2}
\end{figure}

In order to obtain the results of T-odd
functions, which vanish in the lowest order, we need to take the interference between the lowest-order amplitude $M^{(0)}$ and the one-loop-order amplitude $M^{(1)}$
into consideration~\cite{BRODSKY200299,JI200266,COLLINS200243}. The form of $M^{(1)}$ is as follows (see the lower-left part of Fig.~\ref{fig2}, and eikonal approximation~\cite{COLLINS1982445,COLLINS1981381} is already used):
\begin{equation}
    \mathcal{M}^{(1)}(S)=
\begin{cases}
    \int\frac{d^4l}{(2\pi)^4}\frac{\bm{i}e_q\Gamma_s^+(\slashed{p}-\slashed{l}+m)\Upsilon_sU(P,S)}{-(D_1+\bm{i}\varepsilon)(D_2-\bm{i}\varepsilon)(D_3+\bm{i}\varepsilon)(D_4+\bm{i}\varepsilon)}&\text{scalar diquark},\\
    \int\frac{d^4l}{(2\pi)^4}\frac{\bm{i}e_q\varepsilon^*_{\sigma}(P-p,\lambda_a))\Gamma_{a,+}^{\nu\sigma}(\slashed{p}-\slashed{l}+m)d_{\mu\nu}(P-p+l)\Upsilon_a^\mu U(P,S)}{-(D_1+\bm{i}\varepsilon)(D_2-\bm{i}\varepsilon)(D_3+\bm{i}\varepsilon)(D_4+\bm{i}\varepsilon)}&\text{axial-vector diquark},
\end{cases}
\end{equation}
where $D_i$ with $i=1,2,3,4$ in the denominator are propagators:
$D_1=l^2-m^2_g$,
$D_2=l^+$,
$D_3=(p-l)^2-m^2$,
$D_4=(P-p+l)^2-M_X^2$,
and the vertex between the gluon and the scalar ($\Gamma_s$) or axial-vector ($\Gamma_a$) diquark has the following form:
\begin{equation}
    \begin{aligned}
        \Gamma_s^{\mu} &= \bm{i}e_s(2P-2p+l)^{\mu},\\
        \Gamma_{a,\mu}^{\nu\sigma} &= -\bm{i}e_a[(2P-2p+l)_{\mu}g^{\nu\sigma}-(P-p+(1+\kappa)l)^{\sigma}g^{\nu}_{\mu}-(P-p-\kappa~l)^{\nu}g^\sigma_\mu],
    \end{aligned}
\end{equation}
here $e_q$ and $e_{s/a}$ denote the color charge of the quark and the scalar/axial-vector diquark. The symbol $\kappa$ here is the diquark anomalous chromomagnetic moment~\cite{Bacchetta:2008af}, and $\kappa=1$ corresponds to the standard pointlike photon-W coupling~\cite{PhysRevD.33.2608}.
As we can see, it does not show up in the final results under the first choice of $d^{\mu\nu}$. However, it shows up in the results of T-odd TMDs under the last choice of $d^{\mu\nu}$ (see Appendix A). Besides, if we want to get the QCD color interaction,
we shall apply the replacement: $e_qe_X\longrightarrow 4\pi C_F\alpha_s$~\cite{BRODSKY2002344,BRODSKY200299}.

We can obtain TMDs by taking traces between the correlator and some $\gamma$ matrices~\cite{Bacchetta_2007}. Here we introduce
a symbol $\Phi^{[\varGamma]}\equiv\frac{1}{2}\text{Tr}[\Phi\varGamma]$, where $\varGamma$ denotes any possible combination of gamma matrices. We take the Jaffe-Ji-Mulders convention to label TMDs~\cite{PhysRevD.57.5780,MULDERS1996197,PhysRevLett.67.552}.
For the twist-2 case, there are eight TMDs, six for T-evens and two for T-odds. The decomposition expressions of these functions can be found in e.g. Refs.~\cite{MULDERS1996197,PhysRevD.57.5780}.
For the twist-3 case, there are sixteen TMDs, eight for T-evens and eight for T-odds~\cite{Goeke:2005hb,Bacchetta_2007}.
As for the twist-4 case, it has similar forms as the twist-2 case~\cite{Goeke:2005hb} (notice :~$p_T\cdot S_T=-\bm{p}_T\cdot\bm{S}_T$),
\begin{align}
    \Phi^{[\gamma^-]}&=\frac{M^2}{(P^+)^2}[f_3-\frac{\epsilon_T^{ij}\bm{p}_{Ti}\bm{S}_{Tj}}{M}{\color{red}f_{3T}^{\perp}}],\\
    \Phi^{[\gamma^-\gamma_5]}&=\frac{M^2}{(P^+)^2}[\lambda g_{3L}+\frac{\bm{p}_T\cdot\bm{S}_T}{M}g_{3T}],\\
    \Phi^{[\bm{i}\sigma^{i-}\gamma_5]}&=\frac{M^2}{(P^+)^2}[\bm{S}_{T}^ih_{3T}+\frac{\bm{p}_{T}^i}{M}(\lambda h_{3L}^{\perp}+\frac{\bm{p}_T\cdot\bm{S}_T}{M}h_{3T}^{\perp})-\frac{\epsilon_T^{ij}\bm{p}_{Tj}}{M}{\color{red}h_3^{\perp}}],\\
                                      &=\frac{M^2}{(P^+)^2}[\bm{S}_{T}^ih_3+\lambda\frac{\bm{p}_{T}^i}{M}h_{3L}^{\perp}+\frac{(\bm{p}_T^i\bm{p}_{T}^j-\frac{1}{2}\bm{p}_T^2g_T^{ij})\bm{S}_{Tj}}{M^2}h_{3T}^{\perp}-\frac{\epsilon_T^{ij}\bm{p}_{Tj}}{M}{\color{red}h_3^{\perp}}],
\end{align}
where $h_3=h_{3T}+\frac{\bm{p}_T^2}{2M^2}h_{3T}^{\perp}$, and $f_{3T}^{\perp}$ and $h_3^{\perp}$ are T-odd TMDs.\par
Then we can give the unified formulas to calculate T-even and T-odd TMDs, respectively. For the T-even case, we have
\begin{equation}
    \overline{\mathcal{M}}^{(0)}\mathcal{M}^{(0)}=
    \begin{cases}
      \frac{g_s^2(p^2)}{(p^2-m^2)^2}[(\slashed{p}+m\bm{1})\frac{\bm{1}+\gamma_5\slashed{S}}{2}(\slashed{P}+M\bm{1})(\slashed{p}+m\bm{1})] &\text{scalar diquark},\\
      \frac{g_a^2(p^2)}{2(p^2-m^2)^2}[d_{\mu\nu}(\slashed{p}+m\bm{1})\gamma^{\mu}\gamma_5\frac{\bm{1}+\gamma_5\slashed{S}}{2}(\slashed{P}+M\bm{1})\gamma^{\nu}\gamma_5(\slashed{p}+m\bm{1})] &\text{axial-vector diquark}.
    \end{cases}
\end{equation}
By using the definition $\Phi(S)^{[\varGamma]}=\frac{1}{2}\text{Tr}[\Phi(S)\varGamma]$ and Eq.~(\ref{correlator}), we obtain
\begin{equation}
    \label{eqeven}
    \Phi^{[\varGamma]}_{\text{even}}=
    \begin{cases}
      \frac{g_s^2(p^2)}{(p^2-m^2)^2}\frac{(\frac{1}{2}\text{Tr}[\cdots\varGamma])}{(2\pi)^32(1-x)P^+} &\text{scalar diquark},\\
      \frac{g_a^2(p^2)}{(p^2-m^2)^2}\frac{(\frac{1}{2}\text{Tr}[\cdots\varGamma])}{2(2\pi)^32(1-x)P^+} &\text{axial-vector diquark},
    \end{cases}
\end{equation}
here the symbol ``$\cdots$'' denotes the expression inside the square brackets in $\overline{\mathcal{M}}^{(0)}\mathcal{M}^{(0)}$.
Under the dipolar form of $g_X(p^2)$, we have
\begin{equation}
    \label{eqeven2}
    \Phi^{[\varGamma]}_{\text{even}}=
    \begin{cases}
      \frac{g_s^2}{(2\pi)^3}\frac{(1-x)^3}{2[\bm{p}_T^2+L_s^2]^4}\frac{1}{P^+}(\frac{1}{2}\text{Tr}[\cdots\varGamma]) &\text{scalar diquark},\\
      \frac{g_a^2}{2(2\pi)^3}\frac{(1-x)^3}{2[\bm{p}_T^2+L_a^2]^4}\frac{1}{P^+}(\frac{1}{2}\text{Tr}[\cdots\varGamma]) &\text{axial-vector diquark}.
    \end{cases}
\end{equation}
Similarly, for the T-odd case, we can get
\begin{equation}
    \overline{\mathcal{M}}^{(0)}\mathcal{M}^{(1)}=
    \begin{cases}
        \frac{g_s((p-l)^2)[\Gamma_s^+(\slashed{p}-\slashed{l}+m\bm{1})(\bm{1}+\gamma_5\slashed{S})(\slashed{P}+M\bm{1})(\slashed{p}+m\bm{1})/2]}{-A_s(D_1+\bm{i}\varepsilon)(D_2-\bm{i}\varepsilon)(D_3+\bm{i}\varepsilon)(D_4+\bm{i}\varepsilon)} &\text{scalar diquark},\\
        \frac{g_a((p-l)^2)[\Gamma_{a,+}^{\nu\sigma}(\slashed{p}-\slashed{l}+m\bm{1})\gamma^\mu(\bm{1}+\gamma_5\slashed{S})(\slashed{P}-M\bm{1})\gamma^\alpha(\slashed{p}+m\bm{1})d_{\mu\nu}(P-p+l)d_{\sigma\alpha}(P-p)/2]}{-A_a(D_1+\bm{i}\varepsilon)(D_2-\bm{i}\varepsilon)(D_3+\bm{i}\varepsilon)(D_4+\bm{i}\varepsilon)}&\text{axial-vector diquark},
    \end{cases}
\end{equation}
where
\[A_s^{-1}=\frac{e_qg_s(p^2)}{(p^2-m^2)}\!\int\!\frac{d^4l}{(2\pi)^4},~~A_a^{-1}=\frac{e_qg_a(p^2)}{2(p^2-m^2)}\!\int\!\frac{d^4l}{(2\pi)^4}.\]\par
It is a little more complicated because of the loop integral. Firstly, we need to use $\Phi(S)^{[\varGamma]}=\frac{1}{4}\text{Tr}[(\Phi(S)\pm\Phi(-S))\varGamma]+\text{H.c. (abbreviation for Hermitian conjugate)}$ instead
to get our results, and the sign between $\Phi(S)$ and $\Phi(-S)$ depends on the concrete TMD.
The total calculation includes a trace part multiplied with a diagram loop integral part. The trace results of T-odd TMDs are imaginary, so the cut diagram plus its conjugate diagram lead us to take the imaginary part of the loop integral.
Then we can use the Cutkosky rule~\cite{Cutkosky:1960sp} to obtain the imaginary part of the loop integral by putting its intermediate states into on shell:
$\frac{1}{D\pm\bm{i}\varepsilon}\Longrightarrow\mp 2\pi\bm{i}\delta(D)\theta(D^+)$.
Hence, we can get the four-vector of the gluon
\begin{equation}
    l=\left(0,\frac{\bm{l}_T^2-2\bm{l}_T\cdot\bm{p}_T}{2(1-x)P^+},\bm{l}_T\right),
    \label{eql}
\end{equation}
and, similarly, we obtain
\begin{equation}
    \label{eqodd}
    \Phi^{[\varGamma]}_{\text{odd}}=
    \begin{cases}
      \frac{g_s(p^2)}{4(p^2-m^2)}\frac{1}{(2\pi)^3}\frac{e_q}{[2(1-x)P^+]^2}\int\frac{d^2\bm{l}_T}{(2\pi)^2}\frac{\text{Tr}[\cdots\varGamma]g_s((p-l)^2)}{\bm{l}_T^2[(p-l)^2-m^2]} &\text{scalar diquark},\\
      \frac{g_a(p^2)}{4(p^2-m^2)}\frac{1}{2(2\pi)^3}\frac{e_q}{[2(1-x)P^+]^2}\int\frac{d^2\bm{l}_T}{(2\pi)^2}\frac{\text{Tr}[\cdots\varGamma]g_a((p-l)^2)}{\bm{l}_T^2[(p-l)^2-m^2]} &\text{axial-vector diquark},
    \end{cases}
\end{equation}
here the symbol ``$\cdots$'' denotes the expression inside the square brackets in $(\overline{\mathcal{M}}^{(0)}\mathcal{M}^{(1)}(S)\pm\overline{\mathcal{M}}^{(0)}\mathcal{M}^{(1)}(-S))$.
If we choose the dipolar form of $g_X(p^2)$, then we have
\begin{equation}
    \label{eqodd2}
    \Phi^{[\varGamma]}_{\text{odd}}=
    \begin{cases}
      \frac{g_s^2}{4}\frac{1}{(2\pi)^3}\frac{e_q}{4(P^+)^2}\frac{(1-x)^2}{[\bm{p}_T^2+L_s^2]^2}\int\frac{d^2\bm{l}_T}{(2\pi)^2}\frac{\text{Tr}[\cdots\varGamma]}{\bm{l}_T^2[(\bm{p}_T-\bm{l}_T)^2+L_s^2]^2} &\text{scalar diquark},\\
      \frac{g_a^2}{4}\frac{1}{2(2\pi)^3}\frac{e_q}{4(P^+)^2}\frac{(1-x)^2}{[\bm{p}_T^2+L_a^2]^2}\int\frac{d^2\bm{l}_T}{(2\pi)^2}\frac{\text{Tr}[\cdots\varGamma]}{\bm{l}_T^2[(\bm{p}_T-\bm{l}_T)^2+L_a^2]^2} &\text{axial-vector diquark}.
    \end{cases}
\end{equation}\par
Up to now, we establish a framework to calculate all TMDs analytically, and we get the two core formulas, Eq.~(\ref{eqeven}) and Eq.~(\ref{eqodd}).
Then we directly give the results of TMDs, as well as corresponding phenomenological discussions in the following part.\par

For a systematic model calculation, we should also consider the contribution of the one-gluon-exchange final-state interaction for T-even TMDs.
In contrast to T-odd TMDs, we should take the real part, i.e., the principal value term of the loop integral to obtain the results for T-even TMDs,
because their trace results are real.
We list these analytical results in the Appendix.

\section{Analytical results}
\label{sec3}

In this section, we give all the results of TMDs for quark in this spectator model under the light-cone transversely polarized diquark from twist-2 to twist-4.
Namely, we take \[d^{\mu\nu}(P-p)=-g^{\mu\nu}+\frac{(P-p)^{\mu}n_-^{\nu}+(P-p)^{\nu}n_-^{\mu}}{(P-p)\cdot n_-}-\frac{M_a^2}{[(P-p)\cdot n_-]^2}n_-^{\mu}n_-^{\nu}.\]\par

\subsection{Twist-2}
The results of TMDs at twist-2 are the same with the results in Ref.~\cite{Bacchetta:2008af}. It is noted that, under the parametrization of Ref.~\cite{Bacchetta:2008af},
we can find that the Sivers function is negative for u quark and positive for d quark, which is consistent with extractions from experimental data~\cite{Anselmino:2008sga,Collins:2005wb,PhysRevD.85.074008}, as well as calculations on the lattice~\cite{Gockeler:2006zu}.\par

We introduce some symbols to simplify our expressions (``2'' for twist-2, ``e'' for even, ``o'' for odd,  the same below), as
\begin{align}
        \mathcal{A}_{\text{e},s}^2&=\frac{g_s^2}{(2\pi)^3}\frac{(1-x)^3}{2[\bm{p}_T^2+L_s^2]^4},&
        \mathcal{A}_{\text{e},a}^2&=\frac{g_a^2}{(2\pi)^3}\frac{(1-x)}{2[\bm{p}_T^2+L_a^2]^4},\\
        \mathcal{A}_{\text{o},s}^2&=-\frac{g_s^2}{4}\frac{e_qe_s}{(2\pi)^4}\frac{(1-x)^2}{L_s^2[L_s^2+\bm{p}_T^2]^3},&
        \mathcal{A}_{\text{o},a}^2&=-\frac{g_a^2}{4}\frac{e_qe_a}{(2\pi)^4}\frac{(1-x)^2}{L_a^2[L_a^2+\bm{p}_T^2]^3}.
\end{align}

For the T-even case, we have
\begin{align}
    f_1^s&=\mathcal{A}_{\text{e},s}^2[(m+xM)^2+\bm{p}_T^2], & f_1^a&=\mathcal{A}_{\text{e},a}^2[\bm{p}_T^2(1+x^2)+(m+xM)^2(1-x)^2],\\
    g_{1L}^s&=\mathcal{A}_{\text{e},s}^2[(m+xM)^2-\bm{p}_T^2], & g_{1L}^a&=\mathcal{A}_{\text{e},a}^2[\bm{p}_T^2(1+x^2)-(m+xM)^2(1-x)^2],\\
    g_{1T}^s&=\mathcal{A}_{\text{e},s}^2[2M(m+xM)], & g_{1T}^a&=\mathcal{A}_{\text{e},a}^2[2xM(m+xM)(1-x)],\\
    h_{1L}^{\perp s}&=\mathcal{A}_{\text{e},s}^2[-2M(m+xM)], & h_{1L}^{\perp a}&=\mathcal{A}_{\text{e},a}^2[2M(m+xM)(1-x)],\\
    h_{1T}^{\perp s}&=\mathcal{A}_{\text{e},s}^2[-2M^2], & h_{1T}^{\perp a}&=0,\\
    h_1^s&=\mathcal{A}_{\text{e},s}^2[(m+xM)^2], & h_1^a&=\mathcal{A}_{\text{e},a}^2[-2x\bm{p}_T^2].
\end{align}
We notice that $g_{1T}^s=-h_{1L}^{\perp s}=\frac{2M}{m+xM}h_1^s$, and $g_{1T}^a=xh_{1L}^{\perp a}$.\par
Also, we have the following expressions for the T-odd functions:
\begin{align}
    f_{1T}^{\perp s}&=\mathcal{A}_{\text{o},s}^2[M(m+xM)(1-x)], & f_{1T}^{\perp a}&=\mathcal{A}_{\text{o},a}^2[-xM(m+xM)],\\
    h_{1}^{\perp s}&=\mathcal{A}_{\text{o},s}^2[M(m+xM)(1-x)], & h_{1}^{\perp a}&=\mathcal{A}_{\text{o},a}^2[M(m+xM)].
\end{align}
We have the relations $h_1^{\perp s}=f_{1T}^{\perp s}~\text{and}~h_1^{\perp a}=-\frac{1}{x}f_{1T}^{\perp a}$.

\subsection{Twist-3}
More precisely, twist-3 distributions should be calculated from quark-gluon-quark correlators under the collinear expansion framework~\cite{PhysRevD.75.094002}.
However, the calculations of quark-gluon-quark correlations are pretty complicated. We take a rougher but easier way to do our calculations through Dirac decomposition~\cite{Goeke:2005hb,Bacchetta_2007}.
Studies based on the results of TMDs at twist-3 about spin asymmetries, e.g., Refs.~\cite{Mao:2019ibl,Mao:2012dk,Mao:2014aoa,Mao:2016hdi,Lu:2014fva,BACCHETTA2004309}, indicate the feasibility of our approach.
Besides, there are definite relations between quark-quark correlators and quark-gluon-quark correlators, as can be found in Ref.~\cite{Bacchetta_2007}.\par

Our analytical results of twist-3 TMDs are consistent with those in past literatures. Similarly, we introduce the following symbols for simplicity:
\begin{align}
    \mathcal{A}_{\text{e},s}^3&=\frac{g_s^2}{(2\pi)^3}\frac{(1-x)^2}{2[\bm{p}_T^2+L_s^2]^4},&
    \mathcal{A}_{\text{e},a}^3&=\frac{g_a^2}{(2\pi)^3}\frac{(1-x)^2}{2[\bm{p}_T^2+L_a^2]^4},\\
    \mathcal{A}_{\text{o},s}^3&=-\frac{g_s^2}{4}\frac{e_qe_s}{2(2\pi)^4}\frac{(1-x)^2}{L_s^2[L_s^2+\bm{p}_T^2]^3},&
    \mathcal{A}_{\text{o},a}^3&=-\frac{g_a^2}{4}\frac{e_qe_a}{2(2\pi)^4}\frac{(1-x)}{[L_a^2+\bm{p}_T^2]^2}.
\end{align}
So, for the T-even functions, we have
\begin{align}
    e^s&=\mathcal{A}_{\text{e},s}^3[(1-x)(m+M)(m+xM)-(1+\frac{m}{M})\bm{p}_T^2-(x+\frac{m}{M})M_s^2],\\
    e^a&=\mathcal{A}_{\text{e},a}^3[(1-x)(m+M)(m+xM)-(1+\frac{m}{M})\bm{p}_T^2-(x+\frac{m}{M})M_a^2+\frac{2m\bm{p}_T^2}{M(1-x)}],\\
    f^{\perp s}&=\mathcal{A}_{\text{e},s}^3[2mM(1-x)+M^2(1-x^2)-\bm{p}_T^2-M_s^2],\\
    f^{\perp a}&=\mathcal{A}_{\text{e},a}^3[2mM(1-x)+M^2(1+x-x^2)-m^2+\frac{x\bm{p}_T^2-M_a^2}{(1-x)}],\\
    g_L^{\perp s}&=\mathcal{A}_{\text{e},s}^3[\bm{p}_T^2+M_s^2-(1-x)^2M^2],\\
    g_L^{\perp a}&=\mathcal{A}_{\text{e},a}^3[(m+xM)^2+(1-x)M^2+x\bm{p}_T^2-M_a^2],\\
    g_T^{\perp s}&=\mathcal{A}_{\text{e},s}^3[2M^2(1-x)],\\
    g_T^{\perp a}&=\mathcal{A}_{\text{e},a}^3[2M(m+xM)],\\
    g_T^s&=\mathcal{A}_{\text{e},s}^3[(1-x)(m+M)(m+xM)-(x+\frac{m}{M})(\bm{p}_T^2+M_s^2)],\\
    g_T^a&=\mathcal{A}_{\text{e},a}^3[x-\frac{m(1+x)}{M(1-x)}]\bm{p}_T^2,\\
    h_T^{\perp s}&=\mathcal{A}_{\text{e},s}^3[(1-x)(M^2+2mM+xM^2)-\bm{p}_T^2-M_s^2],\\
    h_T^{\perp a}&=\mathcal{A}_{\text{e},a}^3[(m^2-xM^2)+\frac{xM_a^2-\bm{p}_T^2}{(1-x)}],\\
    h_L^s&=\mathcal{A}_{\text{e},s}^3[(1-x)(m+M)(m+xM)+(1-2x-\frac{m}{M})\bm{p}_T^2-(x+\frac{m}{M})M_s^2],\\
    h_L^a&=\mathcal{A}_{\text{e},a}^3[-(1-x)(m+M)(m+xM)+(x+\frac{m}{M})M_a^2-(1-\frac{m(1+x)}{M(1-x)})\bm{p}_T^2],\\
    h_T^s&=\mathcal{A}_{\text{e},s}^3[(1-x)^2M^2-\bm{p}_T^2-M_s^2],\\
    h_T^a&=\mathcal{A}_{\text{e},a}^3[(m^2+2xmM+xM^2)+\frac{\bm{p}_T^2-xM_a^2}{(1-x)}].
\end{align}

While for the T-odd ones, we obtain
\begin{align}
    e_T^{\perp s}={}&\mathcal{A}_{\text{o},s}^3[(1-x)^2M^2-L_s^2-M_s^2],\\
    e_T^{\perp a}={}&\mathcal{A}_{\text{o},a}^3[\frac{xM_a^2-(1-x)(m^2+2xmM+xM^2)-L_a^2}{L_a^2(L_a^2+\bm{p}_T^2)}+\frac{1}{\bm{p}_T^2}\log\left(\frac{L_a^2+\bm{p}_T^2}{L_a^2}\right)],\\
    e_L^s={}&\mathcal{A}_{\text{o},s}^3[(x+\frac{m}{M})(L_s^2-\bm{p}_T^2)],\\
    e_L^a={}&0,
\end{align}
\begin{align}
    e_T^s={}&\mathcal{A}_{\text{o},s}^3[L_s^2+M_s^2-(1-x)(M^2+2mM+xM^2)],\\
    e_T^a={}&\mathcal{A}_{\text{o},a}^3[\frac{(1-x)(m^2-xM^2)-L_a^2+xM_a^2}{L_a^2(L_a^2+\bm{p}_T^2)}+\frac{1}{\bm{p}_T^2}\log\left(\frac{L_a^2+\bm{p}_T^2}{L_a^2}\right)],\\
    f_T^s={}&\mathcal{A}_{\text{o},s}^3[(x+\frac{m}{M})(\bm{p}_T^2-L_s^2)],\\
    f_T^a={}&0,\\
    f_L^{\perp s}={}&\mathcal{A}_{\text{o},s}^3[L_s^2+M_s^2-(1-x)M(2m+M+xM)],\\
    f_L^{\perp a}={}&\mathcal{A}_{\text{o},a}^3[\frac{(1-x)^2M(2m+M+xM)-(1-x+\frac{m}{M})\bm{p}_T^2+(2x+\frac{m}{M})L_a^2-(1-x)M_a^2}{L_a^2(L_a^2+\bm{p}_T^2)}\notag\\
                  {}&~\qquad-\frac{x}{\bm{p}_T^2}\log\left(\frac{L_a^2+\bm{p}_T^2}{L_a^2}\right)],\\
    f_T^{\perp s}={}&0,\\
    f_T^{\perp a}={}&\mathcal{A}_{\text{o},a}^3[\frac{2(L_a^2-\bm{p}_T^2)}{L_a^2(L_a^2+\bm{p}_T^2)}-\frac{1}{\bm{p}_T^2}\log\left(\frac{L_a^2+\bm{p}_T^2}{L_a^2}\right)]\frac{M(m+xM)(1-x)}{\bm{p}_T^2},\\
    g^{\perp s}={}&\mathcal{A}_{\text{o},s}^3[(1-x)^2M^2-L_s^2-M_s^2],\\
    g^{\perp a}={}&\mathcal{A}_{\text{o},a}^3[\frac{(1-x)(m+xM)^2+(1-x)^2M^2+xL_a^2-M_a^2}{L_a^2(L_a^2+\bm{p}_T^2)}-\frac{x}{\bm{p}_T^2}\log\left(\frac{L_a^2+\bm{p}_T^2}{L_a^2}\right)],\\
    h^s={}&\mathcal{A}_{\text{o},s}^3[(x+\frac{m}{M})(\bm{p}_T^2-L_s^2)],\\
    h^a={}&0.
\end{align}\par
Some T-odd TMDs are found to vanish in our model calculation, and most of them can be considered to be relevant with QCD constraints. Due to the time-reversal invariance of
QCD~\cite{COLLINS1993161}, all T-odd TMDs must vanish when integrated upon $\bm{p}_T$, and this implies the following constraints~\cite{Bacchetta_2007,LU2012451,Goeke:2005hb}:
\begin{equation}
    \int d^2\bm{p}_T~f_T(x,\bm{p}_T^2) = 0,~~\int d^2\bm{p}_T~e_L(x,\bm{p}_T^2) = 0,~~\int d^2\bm{p}_T~h(x,\bm{p}_T^2) = 0.
\end{equation}
Although our calculation is model dependent, we can easily find that our results satisfy these constraints.\par
As mentioned, some of these TMDs can offer contributions to azimuthal spin asymmetries by contributing to the structure functions~\cite{Yang:2018aue,Bacchetta_2007,Mao:2013waa,Mao:2014fma}.
The relations between analytical results of TMDs and structure functions can be found in e.g., Ref.~\cite{Bacchetta_2007}, and the structure functions can be expressed as the
combination of TMDs and fragmentation functions. Numerical results based on these analytical results of TMDs can be found in e.g., Refs.~\cite{Mao:2013waa,Mao:2014fma,Mao:2019ibl,Mao:2012dk,Mao:2014aoa},
and relevant physical observables are studied.

\subsection{Twist-4}
For simplicity, similarly, we introduce these following symbols:
\begin{align}
    \mathcal{A}_{\text{e},s}^4&=\frac{g_s^2}{(2\pi)^3}\frac{(1-x)}{4[\bm{p}_T^2+L_s^2]^4},&
    \mathcal{A}_{\text{e},a}^4&=\frac{g_a^2}{(2\pi)^3}\frac{(1-x)}{4[\bm{p}_T^2+L_a^2]^4},\\
    \mathcal{A}_{\text{o},s}^4&=-\frac{g_s^2}{4}\frac{e_qe_s}{2(2\pi)^4}\frac{(1-x)^2}{L_s^2[L_s^2+\bm{p}_T^2]^3},&
    \mathcal{A}_{\text{o},a}^4&=-\frac{g_a^2}{4}\frac{e_qe_a}{2(2\pi)^4}\frac{(1-x)}{[L_a^2+\bm{p}_T^2]^2}.
\end{align}
For the T-even case, we have
\begin{align}
    f_3^s={}&\mathcal{A}_{\text{e},s}^4\{(m+M)^2(1-x)^2-[(1+x+2\frac{m}{M})\bm{p}_T^2+2(1+\frac{m}{M})M_s^2](1-x)+(\frac{\bm{p}_T^2+M_s^2}{M})^2\},\\
    f_3^a={}&\mathcal{A}_{\text{e},a}^4\{(m+M)^2(1-x)^2+[2\frac{m}{M}(\frac{m}{M}-1+x)+(1-x)^2]\bm{p}_T^2-2(1-x)(1+\frac{m}{M})M_a^2+\frac{\bm{p}_T^4+M_a^4}{M^2}\},\\
    g_{3L}^s={}&\mathcal{A}_{\text{e},s}^4\{-(m+M)^2(1-x)^2+[(3-x+2\frac{m}{M})\bm{p}_T^2+2(1+\frac{m}{M})M_s^2](1-x)-(\frac{\bm{p}_T^2+M_s^2}{M})^2\},\\
    g_{3L}^a={}&\mathcal{A}_{\text{e},a}^4\{(m+M)^2(1-x)^2-[2\frac{m}{M}(\frac{m}{M}-1+x)+(1-x)^2]\bm{p}_T^2-2(1-x)(1+\frac{m}{M})M_a^2+\frac{\bm{p}_T^4+M_a^4}{M^2}\},\\
    g_{3T}^s={}&\mathcal{A}_{\text{e},s}^42(1-x)[M(m+M)(1-x)-\bm{p}_T^2-M_s^2],\\
    g_{3T}^a={}&\mathcal{A}_{\text{e},a}^42[(1-x-\frac{m}{M})\bm{p}_T^2+\frac{m}{M}M_a^2-m(m+M)(1-x)],\\
    h_{3T}^s={}&\mathcal{A}_{\text{e},s}^4\{(m+M)^2(1-x)^2-[(1+x+2\frac{m}{M})\bm{p}_T^2+2(1+\frac{m}{M})M_s^2](1-x)+(\frac{\bm{p}_T^2+M_s^2}{M})^2\},\\
    h_{3T}^a={}&\mathcal{A}_{\text{e},a}^42[M_a^2-m^2-M^2(1-x)]\frac{\bm{p}_T^2}{M^2},\\
    h_{3L}^{\perp s}={}&\mathcal{A}_{\text{e},s}^42(1-x)[M(m+M)(1-x)-\bm{p}_T^2-M_s^2],\\
    h_{3L}^{\perp a}={}&\mathcal{A}_{\text{e},a}^42\{[m^2+xM(m+M)-M^2](1-x)+\frac{m}{M}\bm{p}_T^2+(1-x-\frac{m}{M})M_a^2\},\\
    h_{3T}^{\perp s}={}&\mathcal{A}_{\text{e},s}^4[-2M^2(1-x)^2],\\
    h_{3T}^{\perp a}={}&\mathcal{A}_{\text{e},a}^44[M(m+M)(1-x)-M_a^2].
\end{align}
Now we find $g_{3T}^s=h_{3L}^{\perp s}$.\par
For the T-odd case, we have
\begin{align}
    f_{3T}^{\perp s}={}&\mathcal{A}_{\text{o},s}^4[L_s^2+M_s^2-M(m+M)(1-x)],\\
    f_{3T}^{\perp a}={}&\mathcal{A}_{\text{o},a}^4[\frac{\frac{m}{M}M_a^2+(1-x+\frac{m}{M})L_a^2-m(m+M)(1-x)}{L_a^2(L_a^2+\bm{p}_T^2)}+\frac{m}{M\bm{p}_T^2}\log\left(\frac{L_a^2+\bm{p}_T^2}{L_a^2}\right)],\\
    h_{3}^{\perp s}={}&\mathcal{A}_{\text{o},s}^4[L_s^2+M_s^2-M(m+M)(1-x)],\\
    h_{3}^{\perp a}={}&\mathcal{A}_{\text{o},a}^4[\frac{(1-x-\frac{m}{M})M_a^2+\frac{m}{M}L_a^2-(m+M)(M-xM-m)(1-x)}{L_a^2(L_a^2+\bm{p}_T^2)}-\frac{m}{M\bm{p}_T^2}\log\left(\frac{L_a^2+\bm{p}_T^2}{L_a^2}\right)].
\end{align}
Hence, we find $f_{3T}^{\perp s}=h_3^{\perp s}$.\par
TMDs at twist-4 have not been completely calculated yet in earlier papers, and we try to obtain them through our framework. Although there is still a lack of factorization proofs,
the twist-4 TMDs can offer contributions to some physical observables like azimuthal asymmetries, if we expand the cross sections or structure functions
of some process like SIDIS to twist-4 under the framework of collinear expansion. An example for the Sivers asymmetry is showed in next section. Besides,
twist-4 contributions to some physical processes like the SIDIS process~\cite{PhysRevD.83.054010,PhysRevD.95.074017} and semi-inclusive $e^+e^-$ annihilation
process~\cite{PhysRevD.96.054016} are important for further understanding of the feature of hadrons, such as distribution functions and fragmentation functions.\par

In general, all these leading and higher twist TMDs offer us an important opportunity to examine the applicability and effectiveness of TMD factorization~\cite{COLLINS1982445,collins_2011,Ji:2004wu} together with potential extension of factorization proofs to higher twist levels.
We can also get deeper understanding of the nonperturbative properties of QCD, as well as the nontrivial internal structure of hadrons through these TMDs.

\section{Modification on Sivers asymmetry}
\label{sec4}

\begin{figure}[H]
    \centering
    \includegraphics[scale=0.35]{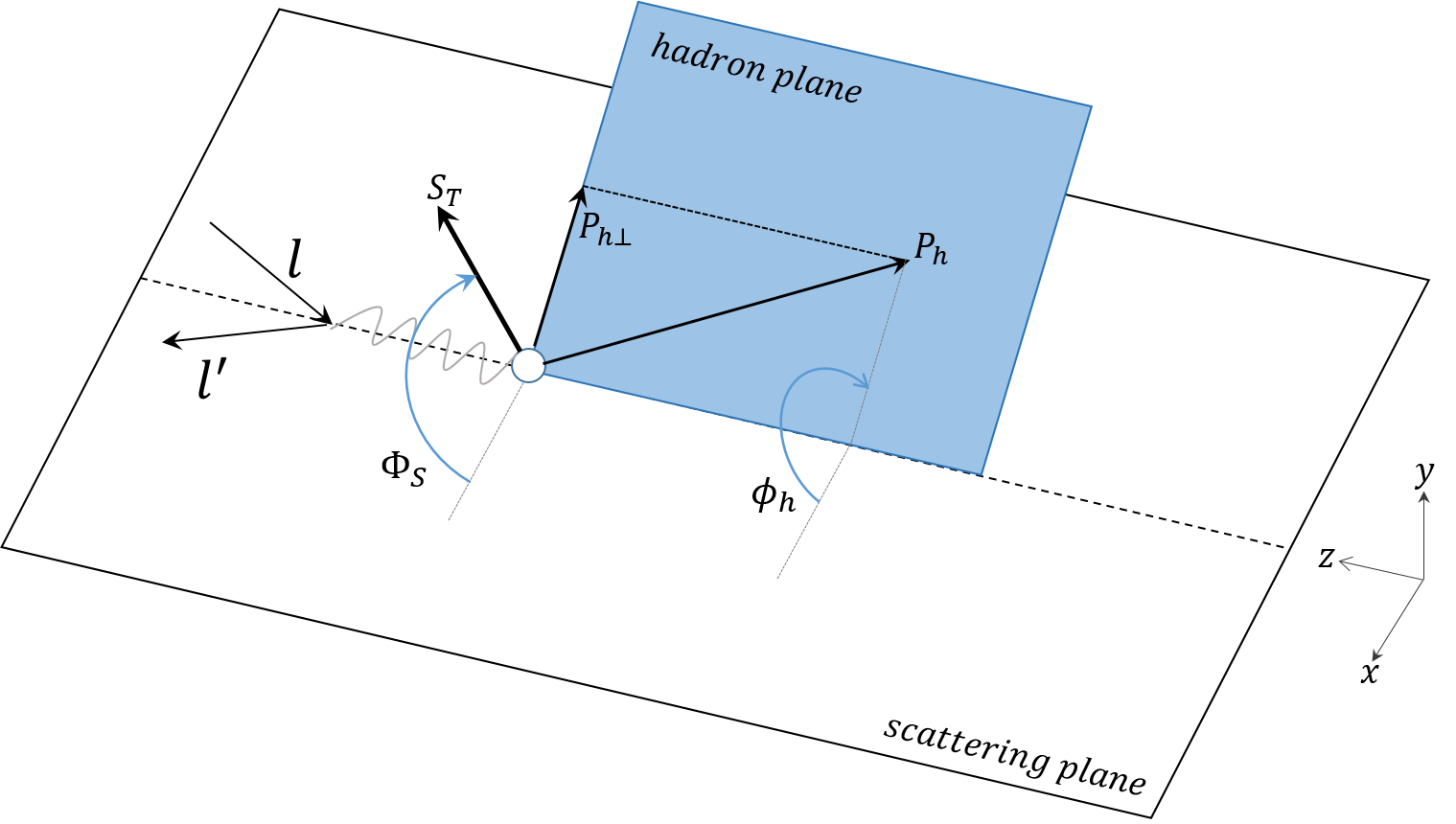}
    \caption{Schematic view of the azimuthal angles in the SIDIS process.}
    \label{fig3}
\end{figure}

In the SIDIS process like $\ell(l)+N(P)\rightarrow\ell(l^\prime)+h(P_h)+X$, the Sivers asymmetry $A_{\mathrm{Siv}}$ (also denoted as $A_{UT}^{\sin(\phi_h-\phi_S)}$ or 2$\langle\sin(\phi_h-\phi_S)\rangle_{UT}$)
is the amplitude of the $\sin(\phi_h-\phi_S)$ modulation
in the distribution of the produced hadrons. Here $\phi_h$ and $\phi_S$ are the azimuthal angles of hadron transverse momentum and the nucleon spin vector,
respectively, in a reference system in which the $z$ axis is the virtual photon direction and the $x$-$z$ plane is the lepton scattering plane (see Fig.~\ref{fig3}), in agreement with the Trento conventions~\cite{Bacchetta:2004jz}.
Under this convention, at leading twist $A_{\mathrm{Siv}}$ can be written as~\cite{Bacchetta_2007,PhysRevLett.103.152002,PhysRevD.85.074008} (notice the sign difference with, e.g., Refs.~\cite{201934,PhysRevD.95.094024})
\begin{equation}
    \label{Asiv} A_{\mathrm{Siv}}(x,z)=-\frac{\sum_qe^2_q~x\int d^2\bm{P}_{h\perp}\mathcal{C}\left[\frac{\bm{P}_{h\perp}\cdot\bm{p}_T}{M|\bm{P}_{h\perp}|}~f_{1T}^{\perp q}(x,\bm{p}_T^2)D_1^q(z,\bm{k}_T^2)\right]}{\sum_qe^2_q~x\int d^2\bm{P}_{h\perp}\mathcal{C}\left[f_1^q(x,\bm{p}_T^2)D_1^q(z,\bm{k}_T^2)\right]}.
\end{equation}
Here the sum is over all (anti)quark flavors, $e_q$ is the quark charge, $D_1^q(z,\bm{k}_T^2)$ is the transverse momentum
dependent unpolarized fragmentation function, and the symbol $\mathcal{C}$ denotes the transverse momentum convolution,
\begin{equation}
    \mathcal{C}[fD]=\int d^2\bm{p}_Td^2\bm{k}_T\delta^{(2)}(z\bm{p}_T+\bm{k}_T-\bm{P}_{h\perp})f(x,\bm{p}_T^2)D(z,\bm{k}_T^2),
\end{equation}
where $\bm{k}_T$ is the transverse momentum of the hadron with respect to the direction of the fragmenting quark, $\bm{P}_{h\perp}$ is the measurable transverse momentum of the
produced hadron, and $z=P^-_h/k^-$ is the longitudinal momentum fraction. The denominator of Eq.~(\ref{Asiv}) can be recast as the collinear expression
\begin{equation}
    \sum_qe^2_q~x\int d^2\bm{P}_{h\perp}\mathcal{C}\left[f_1^q(x,\bm{p}_T^2)D_1^q(z,\bm{k}_T^2)\right]=\sum_qe^2_q~x~f_1^q(x)D_1^q(z).
\end{equation}

Previous studies on Sivers asymmetry only consider the contribution from the leading twist TMD, i.e., the Sivers function $f_{1T}^\perp$.
However, as indicated by Ref.~\cite{PhysRevD.95.074017}, twist-4 TMDs can also participate in this asymmetry provided the cross section or the structure
function is expanded up to twist-4. To utilize our analytical results of twist-4 TMDs, we make an approximation of the Sivers asymmetry formula
by neglecting contributions from the two gluon scattering. Thus, we obtain
\begin{equation}
    A_{\mathrm{Siv}}(x,z)=-\frac{\sum_qe^2_q~x\int dy\int d^2\bm{P}_{h\perp}\mathcal{C}\left[\frac{\bm{P}_{h\perp}\cdot\bm{p}_T}{M|\bm{P}_{h\perp}|}~[f_{1T}^{\perp q}(x,\bm{p}_T^2)+f(y)8x^2(\frac{M}{Q})^2f_{3T}^{\perp q}(x,\bm{p}_T^2)]D_1^q(z,\bm{k}_T^2)\right]}{\sum_qe^2_q~x\int dy~[f_1^q(x)+f(y)8x^2(\frac{M}{Q})^2f_3^q(x)]D_1^q(z)},
\end{equation}
where $f(y)=2(1-y)/[1+(1-y)^2]$ with $y$ the fractional energy of the virtual photon~\cite{Bacchetta_2007}.
This modified formula for the Sivers asymmetry can show us the contribution of twist-4 TMDs intuitively and provide us a tool to further study the
hadronic structures and high twist effects.

\section{Conclusions}
As is shown, we adopt an analytical framework for the calculations of TMDs under the quark-diquark spectator model. Through this framework, we obtain
all TMDs from twist-2 to twist-4 under two types of polarized diquark states. T-even TMDs at one-loop level are also analytically calculated.
Besides, we can also calculate the transverse momentum dependent fragmentation functions in a similar way~\cite{Yang:2018aue}.
The contribution of twist-4 TMDs in the Sivers asymmetry is shown, and a modified formula for this asymmetry is obtained.
Those formulas with model input of various coupling factors can be used to calculate physical observables in an explicit physical process,
so that we can confront theoretical predictions with experimental data for a better understanding of the three-dimensional picture of hadrons.

\section*{Acknowledgements}
This work is supported by the National Natural Science
Foundation of China (Grants No.~12075003, 11605297, 11905187, and 11847217), by the High-level Talents
Research and Startup Foundation Projects for Doctors of Zhoukou Normal University (ZKNUC2016014), and by the Technical Services Project (Grant No.~2019044).
X. W. is supported by China Postdoctoral Science Foundation under Grant No.~2018M640680 and the Academic Improvement Project of Zhengzhou University.

\newpage

\begin{appendices}
    \label{app}

\section{TMDs in another type of polarized diquark}
Here are the results of TMDs from twist-2 to twist-4 under the last choice of $d^{\mu\nu}$ in Eq.~(\ref{eqd}), i.e.,
\[d^{\mu\nu}(P-p)=-g^{\mu\nu}.\]
We make use of the $\mathcal{A}$ symbols above to simplify our results. It is noted that the diquark anomalous chromomagnetic moment $\kappa$ appears
in T-odd TMDs under this kind of $d^{\mu\nu}$.

\subsection{Twist-2}
The results of TMDs at twist-2 are the same with those in Ref.~\cite{Bacchetta:2008af}. For the T-even TMDs, we have
\begin{align}
    f_1^a&=\mathcal{A}_{\text{e},a}^2[(m+xm)^2+2xmM+\bm{p}_T^2](1-x)^2,\\
    g_{1L}^a&=\mathcal{A}_{\text{e},a}^2[(\bm{p}_T^2-m^2-x^2M^2)(1-x)^2],\\
    g_{1T}^a&=\mathcal{A}_{\text{e},a}^2[-2x(1-x)^2M^2],\\
    h_{1L}^{\perp a}&=\mathcal{A}_{\text{e},a}^2[2(1-x)^2mM],\\
    h_{1T}^{\perp a}&=0,\\
    h_1^a&=\mathcal{A}_{\text{e},a}^2[-2x(1-x)^2mM].
\end{align}
For the T-odd TMDs, we have
\begin{align}
    f_{1T}^{\perp a}&=\mathcal{A}_{\text{o},a}^2[\frac{-xM[m(2\kappa+1)+M(2x\kappa+1)]}{2}],\\
    h_1^{\perp a}&=\mathcal{A}_{\text{o},a}^2[\frac{[mM(x(2\kappa-1)+2)+xM^2(2x(\kappa-1)+3)]}{2}].
\end{align}

\subsection{Twist-3}
The results for the T-even case are as follows:
\begin{align}
    e^a&=\mathcal{A}_{e,a}^3[(1-x)(2xM^2+2m^2+mM+xmM)-(2+\frac{m}{M})\bm{p}_T^2-(2x+\frac{m}{M})M_a^2],\\
    f^{\perp a}&=\mathcal{A}_{\text{e},a}^3[(1-x)M(4m+M+xM)-\bm{p}_T^2-M_a^2],\\
    g_L^{\perp a}&=\mathcal{A}_{\text{e},a}^3[(1-x)^2M^2-\bm{p}_T^2-M_a^2],\\
    g_T^{\perp a}&=\mathcal{A}_{\text{e},a}^3[-2(1-x)M^2],\\
    g_T^a&=\mathcal{A}_{\text{e},a}^3[-(1-x)(m^2+xM^2)+x\bm{p}_T^2+xM_a^2],\\
    h_T^{\perp a}&=\mathcal{A}_{\text{e},a}^3[-2(1-x)mM],\\
    h_L^a&=\mathcal{A}_{\text{e},a}^3[-(1-x^2)mM+\frac{m}{M}(\bm{p}_T^2+M_a^2)],\\
    h_T^a&=0.
\end{align}
Here are the T-odd TMDs,
\begin{align}
    e_T^{\perp a}={}&\mathcal{A}_{\text{o},a}^3[\frac{-(1+2\kappa)(1-x)(m^2+2xmM+xM^2)+(1+\kappa)x\bm{p}_T^2+(3+2\kappa)xM_a^2}{2L_a^2(L_a^2+\bm{p}_T^2)}\notag\\
                  {}&~\qquad-\frac{((1+\kappa)(2-x)-1))L_a^2}{2L_a^2(L_a^2+\bm{p}_T^2)}+\frac{(1+\kappa))}{2\bm{p}_T^2}\log(\frac{L_a^2+\bm{p}_T^2}{L_a^2})],\\
    e_L^a={}&\mathcal{A}_{\text{o},a}^3[\frac{m(1-x)}{2M}\frac{(\bm{p}_T^2-L_a^2)}{L_a^2(L_a^2+\bm{p}_T^2)}],\\
    e_T^a={}&\mathcal{A}_{\text{o},a}^3[\frac{(1-x)((1+2\kappa)m^2+2(1-x)mM-(1+2\kappa)xM^2)+(1+\kappa)x\bm{p}_T^2+(3+2\kappa)xM_a^2}{2L_a^2(L_a^2+\bm{p}_T^2)}\notag\\
          {}&~\qquad-\frac{((1+\kappa)(2-x)-1)L_a^2}{2L_a^2(L_a^2+\bm{p}_T^2)}+\frac{(1+\kappa)}{2\bm{p}_T^2}\log(\frac{L_a^2+\bm{p}_T^2}{L_a^2})],\\
    f_L^{\perp a}={}&\mathcal{A}_{\text{o},a}^3[\frac{(1-x)^2M((2+\frac{\kappa}{2})m+M+xM)-(\frac{(3+\kappa)m}{2M}-\kappa x)\bm{p}_T^2-(1+x+\frac{(4+\kappa)m}{2M})M_a^2}{2L_a^2(L_a^2+\bm{p}_T^2)}\notag\\
                  {}&~\qquad-\frac{(1-(1+\kappa)x+\frac{m}{2M})L_a^2}{2L_a^2(L_a^2+\bm{p}_T^2)}-\frac{(1+\kappa)x}{2\bm{p}_T^2}\log(\frac{L_a^2+\bm{p}_T^2}{L_a^2})],\\
    f_T^{\perp a}={}&\mathcal{A}_{\text{o},a}^3[\frac{(L_a^2-\bm{p}_T^2)}{L_a^2(L_a^2+\bm{p}_T^2)}-\frac{1}{2\bm{p}_T^2}\log\left(\frac{L_a^2+\bm{p}_T^2}{L_a^2}\right)]\frac{M(m+xM)(1-x)(1+\kappa)}{\bm{p}_T^2},\\
    f_T^a={}&0,\\
    g^{\perp a}={}&\mathcal{A}_{\text{o},a}^3[\frac{(1-x)[(1+2\kappa)(m+xM)^2+(1-x)(2-x+2\kappa x)M^2]-(1+\kappa)x\bm{p}_T^2-(2+(1+2\kappa)x)M_a^2}{2L_a^2(L_a^2+\bm{p}_T^2)}\notag\\
                {}&~\qquad-\frac{(1-(1-\kappa)x-2\kappa)L_a^2}{2L_a^2(L_a^2+\bm{p}_T^2)}-\frac{(1+\kappa)x}{2\bm{p}_T^2}\log(\frac{L_a^2+\bm{p}_T^2}{L_a^2})],\\
    h^a={}&0.
\end{align}

\subsection{Twist-4}
Similarly, for the T-even functions, we have
\begin{align}
    f_3^a={}&\mathcal{A}_{\text{e},a}^4[(1-x)^2((m+M)^2+2mM)-(1-x)(1+x+4\frac{m}{M})\bm{p}_T^2-2(1-x)(1+2\frac{m}{M})M_a^2 \notag\\
          {}&~\qquad+(\frac{\bm{p}_T^2+M_a^2}{M})^2],\\
    g_{3L}^a={}&\mathcal{A}_{\text{e},a}^4[(1-x)^2(m^2+M^2)-(1-x)(3-x)\bm{p}_T^2-2(1-x)M_a^2+(\frac{\bm{p}_T^2+M_a^2}{M})^2],\\
    g_{3T}^a={}&\mathcal{A}_{\text{e},a}^4[-2(1-x)^2M^2+2(1-x)(\bm{p}_T^2+M_a^2)],\\
    h_{3T}^a={}&\mathcal{A}_{\text{e},a}^4[-2(1-x)^2mM+2(1-x)\frac{m}{M}(\bm{p}_T^2+M_a^2)],\\
    h_{3L}^{\perp a}={}&\mathcal{A}_{\text{e},a}^4[-2(1-x)^2mM],\\
    h_{3T}^{\perp a}={}&0.
\end{align}
For the T-odd functions, we have
\begin{align}
    f_{3T}^{\perp a}={}&\mathcal{A}_{\text{o},a}^4[\frac{-(1-x)((1+2\kappa)(m^2+mM)-(1-x)M^2)+(1+\kappa)\frac{m}{M}\bm{p}_T^2-(1-x-(3+2\kappa)\frac{m}{M})M_a^2}{2L_a^2(L_a^2+\bm{p}_T^2)}\notag\\
                     {}&~\qquad+\frac{(2(1-x)-\frac{m}{M})\kappa L_a^2}{2L_a^2(L_a^2+\bm{p}_T^2)}+\frac{m}{M}\frac{(1+\kappa)}{2\bm{p}_T^2}\log(\frac{L_a^2+\bm{p}_T^2}{L_a^2})],\\
    h_{3T}^{\perp a}={}&\mathcal{A}_{\text{o},a}^4[\frac{(1-x)((1+2\kappa)m^2-(1-2x-2\kappa)mM-3(1-x)M^2)-(1+\kappa)\frac{m}{M}\bm{p}_T^2}{2L_a^2(L_a^2+\bm{p}_T^2)}\notag\\
                     {}&~\qquad+\frac{(3-3x-(3+2\kappa)\frac{m}{M})M_a^2+(2(1-x)(1-\kappa)+\frac{m}{M}\kappa)L_a^2}{2L_a^2(L_a^2+\bm{p}_T^2)}-\frac{m}{M}\frac{(1+\kappa)}{2\bm{p}_T^2}\log(\frac{L_a^2+\bm{p}_T^2}{L_a^2})].
\end{align}

\section{T-even TMDs at one-loop level}
After taking the principal value terms, the loop integral can be rewritten as follows:
\[\int\frac{d^4l}{(2\pi)^4}\frac{1}{(D_1+\bm{i}\varepsilon)(D_2-\bm{i}\varepsilon)(D_3+\bm{i}\varepsilon)(D_4+\bm{i}\varepsilon)}=\int\frac{d^2\bm{l}_T}{(2\pi)^2}\frac{1}{D_1D_3}\int\frac{d~l^+}{(2\pi)}2\,\text{p.v.}(\frac{1}{D_2})\int\frac{d~l^-}{(2\pi)}2\,\text{p.v.}(\frac{1}{D_4}),\]
where $\text{p.v.}(\frac{1}{D}$) denotes the Cauchy principal value of $\frac{1}{D}$.
The Cauchy principal value can be calculated through residue theorem~\cite{e19050215},
\begin{align}
    2\,\text{p.v.}(\frac{1}{D_2})&=2\,\text{p.v.}(\frac{1}{l^+})=2\pi\bm{i}\delta(l^+),\\
    2\,\text{p.v.}(\frac{1}{D_4})&=2\,\text{p.v.}\left[\frac{1}{2(1-x)P^+\left(l^--\frac{\bm{l}_T^2-2\bm{l}_T\cdot\bm{p}_T}{2(1-x)P^+}\right)}\right]=\frac{-2\pi\bm{i}}{2(1-x)P^+}\delta\left(l^--\frac{\bm{l}_T^2-2\bm{l}_T\cdot\bm{p}_T}{2(1-x)P^+}\right).
\end{align}
Thus, we can also get the constraints on $l$, namely, Eq.~(\ref{eql}) still remains.\par
The results of T-even TMDs here only come from the contribution of the one-gluon-exchange final-state interaction, i.e., the gauge link.
The whole results of T-even TMDs from tree level to one-loop level would be the sum of the results above and the results here.

\subsection{Twist-2}
As usual, we introduce some symbols to simplify our expressions,
\begin{align}
    \mathcal{A}_s^2&=\frac{g_s^2}{4}\frac{e_qe_s}{(2\pi)^4}\frac{(1-x)^3}{L_s^2[L_s^2+\bm{p}_T^2]^3},&
    \mathcal{A}_a^2&=\frac{g_a^2}{4}\frac{e_qe_a}{(2\pi)^4}\frac{(1-x)}{L_a^2[L_a^2+\bm{p}_T^2]^3}.
\end{align}
Then we have
\begin{align}
    f_1^s&=\mathcal{A}_s^2[-\bm{p}_T^2], & f_1^a&=\mathcal{A}_a^2[-(1+x^2)\bm{p}_T^2],\\
    g_{1L}^s&=\mathcal{A}_s^2[\bm{p}_T^2], & g_{1L}^a&=\mathcal{A}_a^2[(1+x^2)\bm{p}_T^2],\\
    g_{1T}^s&=\mathcal{A}_s^2[-M(x+xM)], & g_{1T}^a&=\mathcal{A}_a^2[-x(1-x)M(m+xM)],\\
    h_{1L}^{\perp s}&=\mathcal{A}_s^2[M(m+xM)], & h_{1L}^{\perp a}&=\mathcal{A}_a^2[-(1-x)M(m+xM)],\\
    h_{1T}^{\perp s}&=\mathcal{A}_s^2[2M^2], & h_{1T}^{\perp a}&=0,\\
    h_1^s&=0, & h_1^a&=\mathcal{A}_a^2[2x\bm{p}_T^2].
\end{align}

\subsection{Twist-3}
Two symbols for simplicity are as follows:
\begin{align}
    \mathcal{A}_s^3&=\frac{g_s^2}{4}\frac{e_qe_s}{(2\pi)^4}\frac{(1-x)^2}{2L_s^2[L_s^2+\bm{p}_T^2]^3},&
    \mathcal{A}_a^3&=\frac{g_a^2}{4}\frac{e_qe_a}{(2\pi)^4}\frac{(1-x)}{2[L_a^2+\bm{p}_T^2]^2}.
\end{align}
The TMDs are
\begin{align}
    e^s=&\mathcal{A}_s^3[(2-x+\frac{m}{M})\bm{p}_T^2-(x+\frac{m}{M})L_s^2],\\
    e^a=&\mathcal{A}_a^3\frac{2[(1-x)-(1+x)\frac{m}{M}]\bm{p}_T^2}{L_a^2(\bm{p}_T^2+L_a^2)},\\
    f^{\perp s}=&\mathcal{A}_s^3[2\bm{p}_T^2+M_s^2-2mM(1-x)-M^2(1-x^2)-L_s^2],\\
    f^{\perp a}={}&\mathcal{A}_a^3[\frac{(1-x)(1-x+x^2)M^2-2mM(1-x)^2+(1-x)m^2-2x\bm{p}_T^2+M_a^2+xL_a^2}{L_a^2(\bm{p}_T^2+L_a^2)} \notag\\
                {}&\qquad-\frac{x}{\bm{p}_T^2}\log(\frac{L_a^2+\bm{p}_T^2}{L_a^2})],\\
    g_L^{\perp s}=&\mathcal{A}_s^3[(1-x)^2M^2-2\bm{p}_T^2-M_s^2+L_s^2],\\
    g_L^{\perp a}={}&\mathcal{A}_a^3[\frac{(1-x)(1+x-x^2)M^2-2x(1-x)mM-(1-x)m^2-2x\bm{p}_T^2+M_a^2+xL_a^2}{L_a^2(\bm{p}_T^2+L_a^2)} \notag\\
                 {}&\qquad-\frac{x}{\bm{p}_T^2}\log(\frac{L_a^2+\bm{p}_T^2}{L_a^2})],\\
    g_T^{\perp s}=&\mathcal{A}_s^3[-4(1-x)M^2],\\
    g_T^{\perp a}=&\mathcal{A}_a^3[\frac{2(L_a^2-\bm{p}_T^2)}{L_a^2(\bm{p}_T^2+L_a^2)}-\frac{1}{\bm{p}_T^2}\log(\frac{L_a^2+\bm{p}_T^2}{L_a^2})]\frac{(1-x)M(m+xM)}{\bm{p}_T^2},\\
    g_T^s=&\mathcal{A}_s^3[(x+\frac{m}{M})(\bm{p}_T^2-L_s^2)],\\
    g_T^a=&\mathcal{A}_a^3[\frac{(4\frac{m}{M}-3(1-x)(x+\frac{m}{M}))\bm{p}_T^2+(1-x)(m+xM)L_a^2}{L_a^2(\bm{p}_T^2+L_a^2)}+\frac{(1-x)(m+xM)}{\bm{p}_T^2}\log(\frac{L_a^2+\bm{p}_T^2}{L_a^2})]\frac{1}{\bm{p}_T^2},\\
    h_T^{\perp s}=&\mathcal{A}_s^3[2\bm{p}_T^2+M_s^2-2(1-x)mM-(1-x^2)M^2-L_s^2],\\
    h_T^{\perp a}=&\mathcal{A}_a^3[\frac{(1-x)(m^2-xM^2)+2\bm{p}_T^2-xM_a^2-L_a^2}{L_a^2(\bm{p}_T^2+L_a^2)}+\frac{x}{\bm{p}_T^2}\log(\frac{L_a^2+\bm{p}_T^2}{L_a^2})],\\
    h_L^s=&\mathcal{A}_s^3[-(x+\frac{m}{M})(L_s^2+\bm{p}_T^2)-2(1-2x-\frac{m}{M})\bm{p}_T^2],\\
    h_L^a=&\mathcal{A}_a^3\frac{2[1-\frac{m}{M}-(1+x)\frac{m}{M}]\bm{p}_T^2}{L_a^2(\bm{p}_T^2+L_a^2)},\\
    h_T^s=&\mathcal{A}_s^3[-(1-x)^2M^2+2\bm{p}_T^2+M_s^2-L_s^2],\\
    h_T^a={}&\mathcal{A}_a^3[-\frac{(1-x)(m^2+xM^2+2xmM)-xM_a^2+2\bm{p}_T^2-L_a^2}{L_a^2(\bm{p}_T^2+L_a^2)}-\frac{1}{\bm{p}_T^2}\log(\frac{L_a^2+\bm{p}_T^2}{L_a^2})].
\end{align}

\subsection{Twist-4}
Similarly, two symbols for simplicity are
\begin{align}
    \mathcal{A}_s^4&=\frac{g_s^2}{4}\frac{e_qe_s}{(2\pi)^4}\frac{(1-x)}{2L_s^2[L_s^2+\bm{p}_T^2]^3},&
    \mathcal{A}_a^4&=\frac{g_a^2}{4}\frac{e_qe_a}{(2\pi)^4}\frac{(1-x)}{2[L_a^2+\bm{p}_T^2]^2}.
\end{align}
Then, the TMDs are
\begin{align}
    f_3^s=&\mathcal{A}_s^4[((1-x)(x+\frac{m}{M})-\frac{\bm{p}_T^2+M_s^2}{M^2})\bm{p}_T^2-((1-x)(1+\frac{m}{M})-\frac{\bm{p}_T^2+M_s^2}{M^2})L_s^2],\\
    f_3^a=&\mathcal{A}_a^4[\frac{(L_a^2-\bm{p}_T^2-2m^2+2(1-x)mM-(1-x)^2M^2)\bm{p}_T^2}{L_a^2(\bm{p}_T^2+L_a^2)}-\log(\frac{L_a^2+\bm{p}_T^2}{L_a^2})]\frac{1}{M^2},\\
    g_{3L}^s=&\mathcal{A}_s^4[-((1-x)(2-x+\frac{m}{M})-\frac{\bm{p}_T^2+M_s^2}{M^2})\bm{p}_T^2+((1-x)(1+\frac{m}{M})-\frac{\bm{p}_T^2+M_s^2}{M^2})L_s^2],\\
    g_{3L}^a=&\mathcal{A}_a^4[\frac{(L_a^2-\bm{p}_T^2+2m^2-2(1-x)mM+(1-x)^2M^2)\bm{p}_T^2}{L_a^2(\bm{p}_T^2+L_a^2)}-\log(\frac{L_a^2+\bm{p}_T^2}{L_a^2})]\frac{1}{M^2},\\
    g_{3T}^s=&\mathcal{A}_s^4[-(1-x)M(m+M)+2\bm{p}_T^2+M_s^2-L_s^2](1-x),\\
    g_{3T}^a=&\mathcal{A}_a^4[\frac{(1-x)m(m+M)-\frac{m}{M}M_a^2+(1-x-\frac{m}{M})(L_a^2-2\bm{p}_T^2)}{L_a^2(\bm{p}_T^2+L_a^2)}-\frac{(1-x-\frac{m}{M})}{\bm{p}_T^2}\log(\frac{L_a^2+\bm{p}_T^2}{L_a^2})],\\
    h_{3T}^s=&\mathcal{A}_s^4[((1-x)(x+\frac{m}{M})-\frac{\bm{p}_T^2+M_s^2}{M^2})\bm{p}_T^2-((1-x)(1+\frac{m}{M})-\frac{\bm{p}_T^2+M_s^2}{M^2})L_s^2],\\
    h_{3T}^a=&\mathcal{A}_a^4[\frac{(2m^2+(1-x)M(M-m)-M_a^2)\bm{p}_T^2+(M_a^2-(1-x)M(m+M))L_a^2}{L_a^2(\bm{p}_T^2+L_a^2)}+\log(\frac{L_a^2+\bm{p}_T^2}{L_a^2})]\frac{1}{M^2},\\
    h_{3L}^{\perp s}=&\mathcal{A}_s^4[-(1-x)M(m+M)+2\bm{p}_T^2+M_s^2-L_s^2](1-x),\\
    h_{3L}^{\perp a}=&\mathcal{A}_a^4[\frac{-(1-x)(m+M)(m-M+xM)-(1-x-\frac{m}{M})M_a^2+\frac{m}{M}(L_a^2-2\bm{p}_T^2)}{L_a^2(\bm{p}_T^2+L_a^2)}-\frac{m}{M}\frac{1}{\bm{p}_T^2}\log(\frac{L_a^2+\bm{p}_T^2}{L_a^2})],\\
    h_{3T}^{\perp s}=&\mathcal{A}_s^4[2(1-x)M^2],\\
    h_{3T}^{\perp a}={}&\mathcal{A}_a^4\{[\frac{4(M_a^2-(1-x)M(m+M))+2(L_a^2+\bm{p}_T^2)}{L_a^2(\bm{p}_T^2+L_a^2)}-\frac{2}{\bm{p}_T^2}\log(\frac{L_a^2+\bm{p}_T^2}{L_a^2})] \notag\\
                     {}&\qquad+\frac{(1-x)M(m+M)-\bm{p}_T^2-M_a^2}{\bm{p}_T^2}[\frac{2}{L_a^2}-\log(\frac{L_a^2+\bm{p}_T^2}{L_a^2})]\}.
\end{align}

\end{appendices}

\newpage

\bibliographystyle{unsrt}
\bibliography{TMDrefs-PRD-final}

\end{sloppypar}
\end{document}